\documentclass{ws-p8-50x6-00}

\begin{document}


\thispagestyle{empty}

\begin{flushright}
IRB-TH-11/99\\
December, 1999
\end{flushright}

\vspace{2.0cm}

\begin{center}
\Large\bf CURRENT STATUS OF RADIATIVE B DECAYS
\vspace*{0.3truecm}
\end{center}

\vspace{1.8cm}

\begin{center}
\large J. Trampeti\'c\\
{\sl Theoretical Physics Division, Rudjer Bo\v skovi\'c Institute,
\\
P.O.Box 1016, HR-10001 Zagreb, Croatia\\[3pt]
E-mail: {\tt josip@thphys.irb.hr}}
\end{center}
\vspace{1.5cm}

\begin{center}
{\bf Abstract}\\[0.3cm]
\parbox{13cm}
{
A current status report on radiative inclusive and exclusive B decays is presented.
}
\end{center}

\vspace{2.5cm}

\begin{center}
{\sl Talk given at
The $3^{\rm rd}$ International Conference on B Physics and CP Violation,
Taipei, Taiwan, December 3-7, 1999\\
To appear in the Proceedings}
\end{center}


\newpage

\setcounter{page}{1}

\newpage


\title{
CURRENT STATUS OF RADIATIVE B DECAYS}

\author{
JOSIP TRAMPETI\'{C}} 

\address{
Theoretical Physics Division, 
Rudjer Bo\v{s}kovi\'{c} Institute, 
Zagreb, Croatia
\\E-mail: josip@thphys.irb.hr}


\maketitle

\abstracts{
A current status report on radiative inclusive and exclusive B decays is presented.
}

\section{Introduction}

The experimental challenge of finding new physics in direct
search may still take some time if 
new particles or their effects set in only at several hundred GeV. Complementary to 
these direct signals at highest available energies are the measurements of the effects of 
new "heavy" particles in loops, through either precision measurements or detection
of processes occurring only at one loop in the standard model (SM).
Among these are the transitions
induced by flavor-changing neutral currents (FCNC), such as $b \to s \gamma$. 

The first observations of the exclusive $B\to K^* \gamma$ and inclusive
$B\to X_s\gamma$  decays were reported in 1993/94 by the 
CLEO collaboration.\cite{cl}
Skwarnicki from the CLEOII$\&$II.V reported the 
recent exclusive branching ratios:\cite{cl}
\begin{equation}
BR(B^0 \to K^{*0} \gamma) = 4.5 \pm 0.7 \pm 0.3 \times 10^{-5},
\end{equation}
\begin{equation}
BR(B^- \to K^{*-} \gamma) = 3.8 \pm 0.9 \pm 0.3 \times 10^{-5},
\end{equation}
\begin{equation}
BR(B \to K^{*}_2(1430) \gamma) = 1.7 \pm 0.5 \pm 0.1 \times 10^{-5},
\end{equation}
\begin{equation}
BR(B \to K^{*}(1410) \gamma) < 12.7 \times 10^{-5},
\end{equation}
\begin{equation}
BR(B \to K^{*}_2(1430) \gamma)/BR(B \to K^{*} \gamma) = 0.4 \pm 0.1.
\end{equation}
The measurements envolving higher resonant $K^*$-states are still preliminary, 
since the error of $\sim 100 MeV$ in the final-state mass spectrum of means that
we do not know what resonant state has been observed\cite{cl}. 
This year, CLEOII$\&$II.V collaborations have also reported the 
inclusive branching ratio 
\begin{equation}
BR(B\to X_s\gamma) = 3.15 \pm 0.35 \pm 0.41 \times 10^{-4}.
\end{equation}
Using the latest results for inclusive and exclusive branching ratios (BR) , we have obtained
the following central value for the so-called hadronization rate: $R_{K^*}^{exp} \simeq 13\%$.
Note here that Skwarnicki has also reported preliminary results for
exclusive modes based on the quark $b \to d \gamma$ electroweak transition:\cite{cl}\\
$BR(B^0 \rightarrow \rho^0 \gamma) < 4 \times 10^{-5}; 
 BR(B^{0(-)} \rightarrow \omega (\rho^-) \gamma) < 1 \times 10^{-5}$.

\section{NNL QCD Corrected $b\to s \gamma$ Transition}

The $b \rightarrow s \gamma$ process is a one-loop electroweak process,
where the extra gluon exchange (QCD corrections)
changes the nature, i.e. the functional structure of the GIM cancellation:\cite{des}
$(m_t^2-m_c^2)/m_w^2 \rightarrow \ln(m_{t}^2/m_{c}^2)$.

In the standard
model, $B$ decays are described by the effective Hamiltonian obtained
by integrating out the top-quark and $W$-boson fields:
\begin{equation}
H_{\Delta B=1}^{\rm eff}(b \to s \gamma)=
\frac{G_F}{\sqrt{2}}
V_{tb}V_{ts}^*\; \sum_{i=1}^8~c_i(\mu)~O_i(\mu),
\end{equation}
where $c_i$'s are the well-known Wilson coefficients.
The SM theoretical prediction, up to Next-to-Leading Order (NLO)\cite{ciuc} 
in $\alpha_s ln(m_w/m_b)$,
is considerably larger than the lowest-order result.\cite{gr}
Buras et al.\cite{bur} performed a new analysis
by using expansions in powers of $\alpha_s$ and reported a slightly higher,
short-distance (SD) result:
$BR(B \to X_s \gamma )_{\rm NLO}=(3.60\pm 0.33)\times 10^{-4}$.

\section{Exclusive Radiative B Decays}

Exclusive modes are, in principle, affected by large theoretical uncertainties
due to the poor knowledge of nonperturbative dynamics and of a correct 
treatment of large recoil-momenta, determining form factors.

The  Lorentz decomposition of the $B-K^*$  matrix element
of the operator $O_7$, taking into account the gauge condition, the current
conservation, 
the spin symmetry and for the on-shell photon, gives the following
hadronization rate $R_{K^*}$:\cite{excl}
\begin{equation}
R_{K^*}=
\frac{{\Gamma(B\to K^* \gamma)}}{{ \Gamma(b\to s\gamma)}} \\
 = [\frac{m_b(m^2_B-m^2_{K^*})}{m_B(m^2_b-m^2_s)}]^3
(1+\frac{m^2_s}{m^2_b})^{-1} |T_1^{K^*}(0)|^2. 
\end{equation}

Since the first calculation of the hadronization rate
$R_{K^*} \approx 6\%$ by Deshpande at al.\cite{excl}
a large number of papers have reported $R_{K^*}$
from the range of 5\% to (unrealistic) 80\%.
Different methods have been employed, from
simple quark models, QCD sum rules, HQET and chiral symmetry, QCD on the lattice, 
light cone sum rules, to the
perturbative QCD type of evaluations of exclusive modes.\cite{excl}
In any event, the above form factor will be obtained
in the future from first-principle calculations
on the lattice. 
Recently, it looks like that the hadronization rate calculations
have stabilized around 10\%.

If long-distance (LD) and other nonperturbative effects
are neglected, two exclusive modes
are connected by a simple relation
\begin{equation}
BR(B \rightarrow \rho \gamma) =
\xi^{2} \left| V_{td}/V_{ts} \right|^{2}
BR(B \rightarrow K^{*} \gamma),
\end{equation}
where $\xi$ measures the SU(3) breaking effects.
They are typically of the order of 30\%.\cite{excl}
Misiak has reported the SD BR's:\cite{mis} 
\begin{equation}
BR(b \to d \gamma)=1.61\times 10^{-5}; 
BR(B^+ \rightarrow \rho^+ \gamma) =  [1,4] \times 10^{-6},
\end{equation}
\begin{equation}
BR(B^0 \rightarrow \rho^0 \gamma) =
BR(B^0 \rightarrow \omega \gamma) =
[0.5,2] \times 10^{-6}.
\end{equation}

\section{LD Contributions to Inclusive/Exclusive 
B Decays}

First note that LD corrections cannot be computed
from first principles; it is possible to
estimate them phenomenologically.\cite{pak} For instance, the operators
$O_{1,2}$ contain the $\bar cc$ current. So
one could imagine the $\bar cc$ pair propagating through a long distance,
forming intermediate $\bar cc$ states  
(off-shell $J/\psi$'s), which turn into a photon via the vector meson dominance (VMD).

The total (SD + LD) amplitude for $b \to d(s) \gamma$ is
\cite{pak}
\begin{eqnarray}
\lefteqn{M(b\rightarrow d \gamma ) 
= -{eG_F\over 2\sqrt{2}}[V_{tb}V_{td}^* ( {m_b\over 4\pi^2} c^{eff}_7(\mu)
-{2\over 3}a_2\sum_i{g_{\psi_i}^2(0)\over m_{\psi_i}^2 m_b} )}\nonumber\\
& &-{a_2\over m_b}V_{ub}V_{ud}^* ( {2\over 3}\sum_i{g_{\psi_i}^2(0)\over
m_{\psi_i}^2}
-{1\over 2}{g_\rho^2(0)\over m_\rho^2} -{1\over 6} {g_\omega^2(0)
\over m_{\omega}^2})]
\bar d \sigma^{\mu\nu}(1+\gamma_5)b F_{\mu\nu}.
\end{eqnarray}
If in the above equation we replace the d by the s quark
and forget the last three terms, then
we obtain the total amplitude for the $b \to s \gamma$ decay.
Deshpande et al\cite{pak} have found a strong suppression when
extrapolating $g_{\psi}(m_{\psi}^2)$ to $g_{\psi}(0)$: 
$g^2_{\psi(1S)}(0)/g^2_{\psi(1S)}(m^2_{\psi}) = 0.13 \pm 0.04$.
The LD contributions to an inclusive mode and to its exclusive modes
are all found to be small, 
typically  one to two orders of magnitude below the SD's.\cite{pak}

\section{Discussion and Conclusions}

Note that the so-called spectator-quark contributions\cite{don} 
and the first calculable nonperturbative,
essentialy long distance, correction\cite{vol}
to the inclusive rate are of the order of a few percent. It has also been proved that
the fermionic (quarks and leptons) and photonic loop corrections to $b \to s \gamma$ reduce
$BR(b \to s \gamma)/BR(b \to ce\bar \nu)$ by $\sim 8\pm2\%$.\cite{marc}
Consequently, it 
is more appropriate to use $\alpha_{em}=1/137$
for the real photon emission.\cite{marc,mis} 

In general, we can conclude that, in theory, more effort is required in
calculating quark (inclusive) decays through higher loops.
A better understanding of bound states of heavy-light quarks (B-meson ect.) 
and highly recoiled light-quark bound states ($K^*$, $\rho$, \ldots)
is desirable.
This can be achived by inventing new, more sophisticated (perturbative)
methods\cite{mel} and apply them
to the calculation of radiative B-meson decays, which incorporate the full spectrum
of quark bound states ($K^*$, $\rho$, $K_1^*$, \ldots).
In experiment, with a larger amount of data we might expect a
stable but small increase of $BR(b \to s \gamma)$,
a decrease of $BR(B \to K^* \gamma)$, determinations
of $BR(B \to K_1^* \gamma)$ and $BR(B \to K_2^* \gamma)$ and 
first measurements of $BR(b \to d \gamma)$ and $BR(B \to \rho \gamma)$.


\begin{thebibliography}{99}

\bibitem{cl}  R. Ammar {\it et al}, \Journal{\PRL}{71}{674}{1993}; 
M.S. Alam {\it et al},\\ \Journal{\PRL}{74}{2885}{1995};
T. Skwarnicki, this Proceedings.

\bibitem{des} S. Bertolini, F. Borzumati, A. Masiero, \Journal{\PRL}{59}{180}{1987};
N.G. Deshpande, P. Lo, J. Trampeti\' c, G. Eilam, P. Singer,\\
\Journal{\PRL}{59}{183}{1987}.

\bibitem{ciuc} K. Chetyrkin, M. Misiak, M. M$\ddot u$nz, \Journal{\PLB}{400}{206}{1997};\\
C. Greub, T. Hurth, \Journal{PRD}{56}{2934}{1997};
M. Ciuchini, G. Degrassi,\\ P. Gambino, G.F. Giudice, \Journal{\NPB}{527}{21}{1998}.

\bibitem{gr} B. Grinstein, R. Springer, M.B. Wise, 
\Journal{\NPB}{339}{269}{1990};

\bibitem{bur} A. Buras, A. Kwiatkowski, N. Pott, \Journal{\PLB}{414}{157}{1997};\\
\Journal{\PLB}{434}{459}{1998}(erratum).

\bibitem{excl} N.G. Deshpande {\it et al}, \Journal{\PRL}{59}{183}{1987}; 
{\it Z. Phys.} {\bf C40}, 369 (1988).
P. Colangelo, C.A. Dominguez, G. Nardulli, N. Paver, \Journal{\PLB}{317}{183}{1993};
R. Casalbuoni, A. Deandrea, N. Di Bartolomeo, R. Gatto, G. Narduli,
\Journal{\PLB}{312}{315}{1993};
K.C. Bowler {\it et al},  \Journal{\PRL}{72}{1398}{1994};
L. Del Debbio, J. Flynn, L. Lellouch,\\ J. Nieves,  \Journal{\PLB}{416}{392}{1998};
P. Ball, V.M. Braun, \Journal{\PRD}{58}{094016}{1998};
H.H. Asatryan, H.M. Asatryan,\\ D. Wyler, \Journal{\PLB}{470}{223}{1999}.

\bibitem{mis} M. Misiak, hep-ph/0002007, this Proceedings.

\bibitem{pak} E. Golowich, S. Pakvasa,  \Journal{\PLB}{205}{393}{1988};\\ 
J. Trampeti\'c, {\it Fizika} {\bf B2}, 121 (1993); and a paper in preparation;\\  
N.G. Deshpande, X.-G. He, J. Trampeti\' c, \Journal{\PLB}{367}{362}{1996}.

\bibitem{don} J.F. Donoghue, A.A. Petrov \Journal{\PRD}{53}{3664}{1996}.

\bibitem{vol} M.B. Voloshin, \Journal{\PLB}{397}{275}{1997}.

\bibitem{marc} A. Czarnecki, W. Marciano, \Journal{\PRL}{81}{277}{1998};

\bibitem{mel} H. Simma, D. Wyler, \Journal{\PLB}{272}{395}{1991};\\
B. Meli\' c, \Journal{\PRD}{59}{074005}{1999};
M. Benecke, G. Buchalla,\\ M. Neubert, C.T. Sachrajda, \Journal{\PRL}{83}{1914}{1999}.

\end{thebibliography}
\end{document}